\documentclass[a4paper,10pt]{article}

\usepackage[margin=1in]{geometry}

\pdfoutput=1

\usepackage{graphicx}
\usepackage{bm}
\usepackage{dcolumn}
\usepackage{setspace}
\usepackage{abstract}
\usepackage{multirow}
\usepackage{amsmath}
\usepackage{fancyhdr}

\usepackage{cite}
\usepackage{color,soul}

\usepackage[affil-it]{authblk}

\usepackage[utf8]{inputenc}
\usepackage{lineno,hyperref}

\hypersetup{colorlinks=true, linkcolor=blue, citecolor=red}

\begin{document}

\title{\bf Influence of twist boundary on deformation behaviour of $<$100$>$ BCC Fe nanowires}
\date{}

\author{G. Sainath\footnote{email : sg@igcar.gov.in}, B.K. Choudhary\footnote{email : bkc@igcar.gov.in}}

\affil {Deformation and Damage Modelling Section, Mechanical Metallurgy Division \\
Indira Gandhi Centre for Atomic Research, HBNI, Kalpakkam,  
Tamilnadu-603102, India}

\maketitle

\doublespacing

\begin{onecolabstract}

Molecular dynamics simulations revealed significant difference in deformation behaviour of $<$100$>$ 
BCC Fe nanowires with and without twist boundary. The plastic deformation in perfect $<$100$>$ BCC Fe 
nanowire was dominated by twinning and reorientation to $<$110$>$ followed by further deformation by 
slip mode. On the contrary, $<$100$>$ BCC Fe nanowire with a twist boundary deformed by slip at low
plastic strains followed by twinning at high strains and absence of full reorientation. The results 
suggest that the deformation in $<$100$>$ BCC Fe nanowire by dislocation slip is preferred over twinning 
in the presence of initial dislocations or dislocation networks. The results also explain the absence 
of extensive twinning in bulk materials, which inherently contains large number of dislocations.\\

\end{onecolabstract}


{\small 

\section{Introduction}

In recent years, metallic nanowires have attracted a major attention for research due to their unique 
properties and potential applications in future nano/micro electro-mechanical systems (NEMS/MEMS). Due 
to good magnetic properties, BCC Fe nanowires in particular find applications in high-density magnetic 
recording media, data storage and memory devices, spin electronics and smart sensors \cite{Appl-1,Appl-2}. 
The reorientation and the associated pseudo elastic and shape memory behaviour of nanowires originating 
from deformation by twinning facilitate important applications in smart sensors. In view of this, it 
is essential to understand the occurrence of twinning and the factors that influences the twinning 
mode of deformation in metallic nanowires.

Deformation twinning usually occurs under conditions that lead to high stresses such as high strain 
rates or low temperatures \cite{Mahajan}. Twinning in perfect nanowires occurs due to high stresses 
resulting from small size and the exhaustion of dislocation sources \cite{Origin,Source}. Several 
experimental \cite{fcc-exp,bcc-exp} and atomistic simulation studies \cite{Park-fcc,Cai,Li,Sai-CMS16} 
have shown that both FCC and BCC metallic nanowires deform by twinning 
mechanism. Using in-situ scanning electron microscopy and high-resolution transmission electron 
microscopy, it has been shown that the plastic deformation in defect-free Au nanowire with $<$110$>$ 
orientation occurs by twinning mechanism \cite{fcc-exp}. In agreement with experimental studies, 
the atomistic simulations also revealed deformation dominated by twinning in Cu, Au and Ni nanowires 
with $<$110$>$ orientation under tensile loading \cite{Park-fcc,Cai,Liang}. It has also been shown 
that twinning occurs in $<$100$>$ Cu, Au and Ni nanowires under compressive loading \cite{Park-fcc}. 
Similar to FCC nanowires, Wang et al. \cite{bcc-exp} reported the first experimental evidence of
deformation twinning in BCC nanowires. It has been shown that the deformation occurs by twinning under 
tensile loading in  $<$100$>$ orientation and under compressive loading in  $<$110$>$ and  $<$111$>$ 
orientations in BCC W nanowires \cite{bcc-exp}. Molecular dynamics simulations have also shown that 
the deformation proceeds by twinning under tensile loading in  $<$100$>$ BCC Fe, Mo and W nanowires 
\cite{Li,Sai-CMS16,Cao,Mo,Healy}. In both FCC and BCC nanowires, deformation by twinning leads to 
reorientation, pseudo-elasticity, shape memory and super-elasticity \cite{Li,Liang,Cao,Park-PRL,Sutrakar,Shi}. 
Most of these studies were focused on perfect nanowires, and it is not clear whether the deformation 
by twinning will continue to occur in the presence of defects such as dislocations and grain boundaries. 
Recently, it has been shown that the presence of twin boundary in  $<$110$>$ FCC Cu nanopillars changes 
the deformation mechanism from twinning to slip \cite{Sai-PLA}. In the present study, an attempt has 
been made to understand the influence of twist boundary on deformation twinning in BCC Fe nanowires.

\section{MD simulation details}

Molecular dynamics simulations have been carried out using large-scale atomic/molecular massively parallel 
simulator (LAMMPS) package \cite{Plimpton-1995}, and the visualization of atomic structure was accomplished
using AtomEye \cite{J-Li-2003} package. The Burgers vector and the total length of dislocations have been 
determined using OVITO \cite{OVITO}. Initially, single crystal BCC Fe nanowires 
oriented in $<$100$>$ axial direction with \{100\} side surfaces were created. The nanowire had dimensions 
of $8.5\times8.5\times17.1$ nm$^3$ consisting of about 110,000 atoms arranged in a BCC lattice and interacting 
through an embedded atom method (EAM) potential developed by Mendelev et al. \cite{Mendelev-2003}. Following the 
creation of perfect nanowire, the nanowire is divided into two equal upper and lower grains along the nanowire 
axis (Figure \ref{Initial}a). In order to introduce a twist boundary, the upper grain is rotated by an angle 
$+2^o$ and lower grain by $-2^o$ around the nanowire axis. Upon relaxation, a screw dislocation network was 
spontaneously formed at the interface separating upper and lower grains (Figure \ref{Initial}b and c). The 
second network at the top appears due to the periodic boundary conditions. The choice of twist angle of $+2^o$ 
was based on earlier study \cite{Caro-2011} on the formation of initial dislocation network structure in BCC 
Fe. The dislocation network has square structure (four-fold symmetry) with junctions formed by four $<$100$>$ 
type sessile screw dislocations $b_1, b_2, b_3$ and $b_4$ shown in Figure \ref{Initial}c. The Burger vectors 
of dislocations at the junction satisfy the relation $b_1 + b_2 = b_3 + b_4$. Following the creation of \{100\} 
twist boundary, initial velocities were assigned randomly to all the atoms according to finite temperature 
Maxwell distribution and then the system was equilibrated to 10 K in canonical ensemble (constant NVT). The 
velocity verlet algorithm was used to integrate the equation of motion with a time step of 2 fs. Upon completion 
of equilibration process, tensile deformation was carried out at a constant engineering strain rate of 
$1\times10^8$ s$^{-1}$ along the nanowire axis. The average stress is calculated from the Virial expression 
\cite{Virial}.

\begin{figure}[h]
\centering
\includegraphics[width= 10cm]{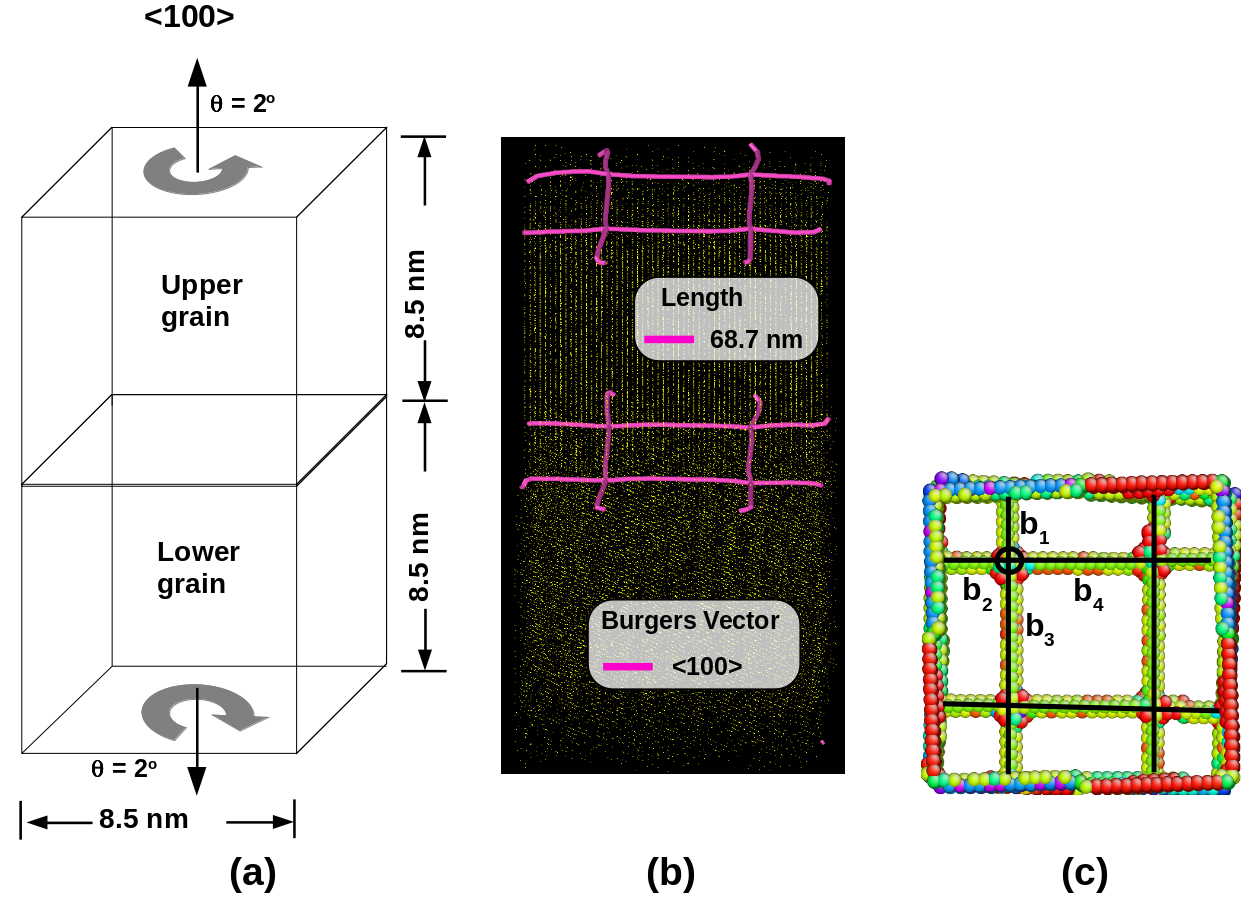}
\caption {\footnotesize Schematic of the process of creating a twist grain boundary in $<$100$>$ BCC Fe 
nanowire. The $<$100$>$ sessile dislocation network formed at the twist boundary (TWB) is shown in (b) 
and (c). In (b), only dislocation network is shown using OVITO. The atoms are coloured according to the 
centro-symmetry parameter in (c).}
\label{Initial}
\end{figure}

\section{Results and discussion}

The stress-strain behaviour of $<$100$>$ BCC Fe nanowire with \{100\} twist boundary along with perfect 
nanowire under tensile loading is shown in Figure \ref{stress-strain}. Both the nanowires displayed similar 
elastic deformation at low strains having elastic modulus of 164 GPa. The Young’s modulus value of 164 GPa 
for perfect  $<$100$>$ BCC Fe nanowire is in good agreement with those obtained using MD simulations 
\cite{Sai-CMS16} and Ab-initio calculations \cite{Friak}. The perfect nanowire exhibited large elastic 
deformation and higher yield strength compared to that displayed by nanowire with a twist boundary. 
The yield strength of 12.4 GPa obtained for perfect nanowire is close to the theoretical strength of BCC 
Fe in $<$100$>$ direction \cite{strength} and the yielding leads to an abrupt large drop in flow stress 
to about 2 GPa. BCC Fe nanowire with \{100\} twist boundary displayed comparatively small elastic deformation
along with lower yield strength of 5.3 GPa and lower strain to yielding ($\varepsilon = 0.032$) compared to 
perfect nanowire. Following the initial 
yielding, the two nanowires exhibited contrasting flow behaviour during plastic deformation. A constant flow 
stress of 2 GPa up to $\varepsilon = 0.63$ followed by a second elastic peak, yield drop and continuous 
decreases in flow stress till failure was observed in the perfect BCC Fe nanowire. Contrary to this, the 
nanowire containing a twist boundary exhibited large flow stress fluctuations at low strains followed by a 
second elastic peak with peak stress value 8.6 GPa and yield drop at $\varepsilon = 0.15$. After second 
yield drop, a constant but marginally lower flow stress of 1.8 GPa up to $\varepsilon = 0.50$ followed by 
decrease in flow stress till failure was observed. It can be seen in Figure \ref{stress-strain} that nanowire 
with twist boundary has a significantly lower strain to failure than perfect nanowire.

\begin{figure}[h]
\centering
 \includegraphics[width = 9cm]{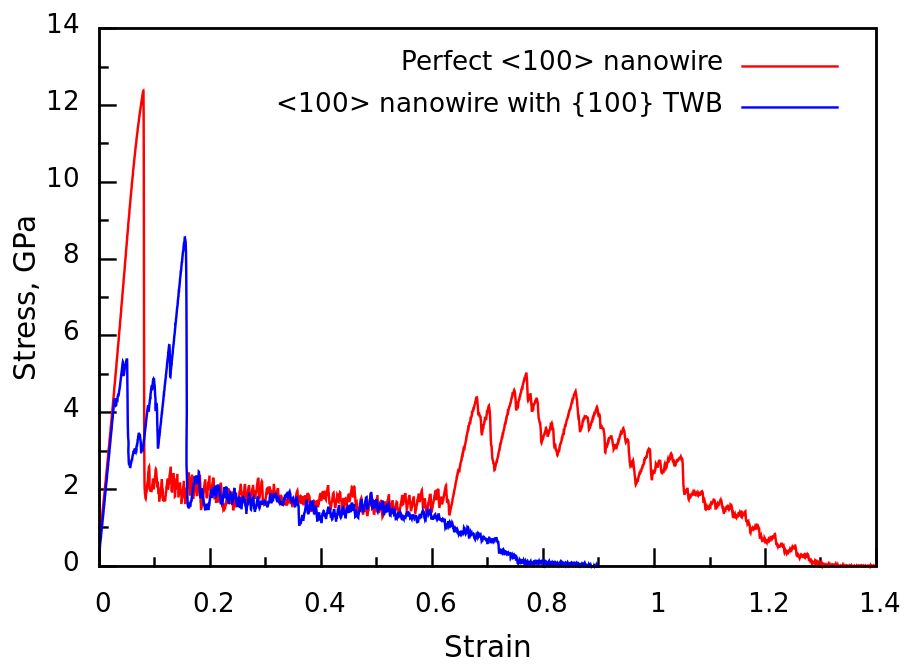}
 \caption {\footnotesize Stress-strain behaviour of  $<$100$>$ BCC Fe perfect nanowire and with twist boundary 
 (TWB).}
 \label{stress-strain}
 \end{figure}
 
The atomic snapshots displaying the deformation behaviour of perfect BCC Fe nanowires at 10 K are shown in 
Figure \ref{perfect}. The nanowire yielded by the nucleation of a twin embryo from the corner on \{112\} 
plane with a twin front propagating in $<$111$>$ direction is seen in Figure \ref{perfect}a. Once the twin 
front reaches the opposite surface, the twin embryo becomes a full twin enclosed by two twin boundaries 
(Figure \ref{perfect}b). Along these twin boundaries, the 1/6$<$111$>$ twinning partial dislocations move 
in opposite directions and displace the twin boundaries away from each other. As a result, the twin grows 
plane by plane along the nanowire axis (Figure \ref{perfect}c). Since twin growth process does not create 
any new surface, the flow stress plateau during the twin growth is observed (Figure \ref{stress-strain}). 
In the absence of obstacles, the twin boundaries easily sweep across the nanowire and due to periodic 
boundary conditions along the length, they meet each other and annihilate. This leads to reorientation of
initial $<$100$>$ nanowire to $<$110$>$ nanowire (Figure \ref{perfect}d) along with a change in cross-section
shape from square (Figure \ref{perfect}a-c) to rhombic (Figure \ref{perfect}d-f). Following reorientation, 
$<$110$>$ nanowire undergoes an elastic deformation at $\varepsilon = 0.63$ with lower modulus (Figure 
\ref{stress-strain}) and yielding by the nucleation of 1/2$<$111$>$ full dislocations is shown in Figure 
\ref{perfect}e. Further plastic deformation in the reoriented nanowire occurs by dislocation slip mechanism 
till failure (Figure \ref{perfect}e-f). Following second yielding, the necking initiates at the strain value 
of 0.95 followed by continuous decrease in flow stress (Figure \ref{stress-strain}) associated with the growth
of necking leading to failure (Figure \ref{perfect}f).

\begin{figure}[h]
\centering
\includegraphics[width = 12cm]{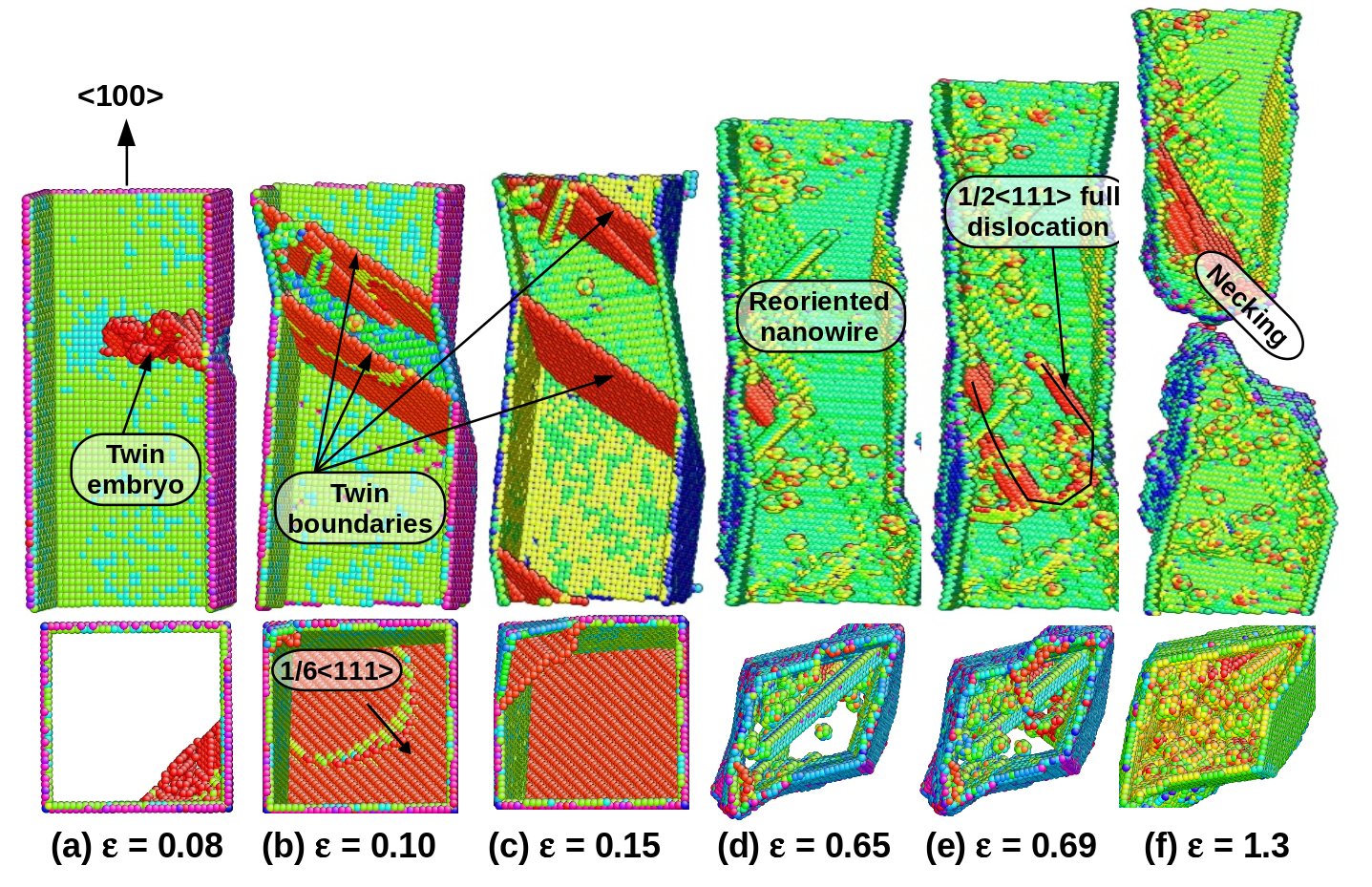}
\caption {\footnotesize Deformation behaviour of perfect $<$100$>$ BCC Fe nanowire under tensile loading. 
The atoms are coloured according to the centro-symmetry parameter. The perfect BCC Fe atoms and surfaces 
were removed for clarity.}
\label{perfect}
\end{figure}
 
The deformation behaviour of BCC Fe nanowire with twist boundary at 10 K is shown in Figure \ref{twist}. 
During elastic deformation, no change in the initial dislocation network structure has been observed. The 
yielding occurs by splitting $<$100$>$ type sessile screw dislocations into two 1/2$<$111$>$ glissile 
dislocations (Figure \ref{twist}a). As a result, the decrease in total length of $<$100$>$ dislocations 
from 68.7 to 64.7 nm at an expense of 1/2$<$111$>$ dislocations is seen (Figure \ref{twist}a). The 
splitting of sessile screw dislocations initiates from the surface of the nanowire and penetrates towards 
the dislocation junction. With increasing plastic strain, more and more sessile dislocations splits into 
glissile dislocations and move away from the initial network (Figure \ref{twist}b). This continuous 
splitting increases the total length of 1/2$<$111$>$ glissile dislocations from 10.1 nm at yielding 
($\varepsilon = 0.032$) to 96.6 nm at $\varepsilon = 0.051$ with corresponding decrease in $<$100$>$ 
sessile dislocations. Energetically, the splitting of $<$100$>$ dislocations is difficult to be observed 
at low stresses due to increase in energy, i.e. $a^2 < 3a^2/4 + 3a^2/4$. However, this dislocation 
reaction becomes feasible at high stresses typically in the order of GPa as observed in the present 
study. The continuous splitting of $<$100$>$ network dislocations followed by glide of resultant dislocations 
and their escape to surface leads to small elastic peaks and flow stress drops at low strains in the 
range 0.032-0.15 (Figure \ref{stress-strain}). The presence of high surface and image stresses aided 
by small size facilitates dislocations to escape from the nanowire. As a result, the total length of 
1/2 $<$111$>$ glissile dislocations also decreases from a peak of 96.6 nm at $\varepsilon = 0.051$ 
(Figure  \ref{twist}b) to 37.5 nm at $\varepsilon = 0.08$ (Figure \ref{twist}c). The continuous split 
and escape of dislocations leave the nanowire in a dislocation-free state (Figure \ref{twist}d) with a 
few point defects. At this stage, an increase in plastic strain leads to second elastic deformation with 
peak stress value of 8.6 GPa followed by an abrupt drop in the flow stress. This abrupt drop due to the 
yielding of dislocation-free nanowire takes place by the nucleation of two-layer twin embryo from the 
slip step as shown in Figure \ref{twist}e. With increasing plastic deformation, the two layer twin embryo 
becomes a full twin enclosed by two twin boundaries as shown in Figure \ref{twist}f. This twin boundaries
move away from each other (Figure \ref{twist}g) by the repeated initiation and glide of 1/6$<$111$>$ 
twinning partial dislocations resulting in the constant flow stress with some oscillations in the strain 
range 0.15-0.6 (Figure \ref{stress-strain}). Due to the presence of point defects generated by the movement 
of 1/2$<$111$>$  dislocations during initial deformation by slip mode, the twin boundary migration or twin 
growth process is impeded and as a result, the nanowire does not undergo full reorientation and fails by 
shearing along the \{112\} twin boundary plane (Figure \ref{twist}h). The strain for the onset of necking 
has been obtained as 0.62. These results indicate that the nanowire containing a twist boundary deforms 
by slip at small strains followed by twinning at large strains. The simulations performed on nanowire 
with multiple twist boundaries with smaller twist boundary spacing also indicated similar deformation
behaviour. Further, MD simulations performed at higher temperatures of 300 and 600 K indicated that the 
perfect nanowire and the nanowire containing a twist boundary undergo deformation similar to that at 10 K.

\begin{figure}[h]
\centering
\includegraphics[width = 10cm]{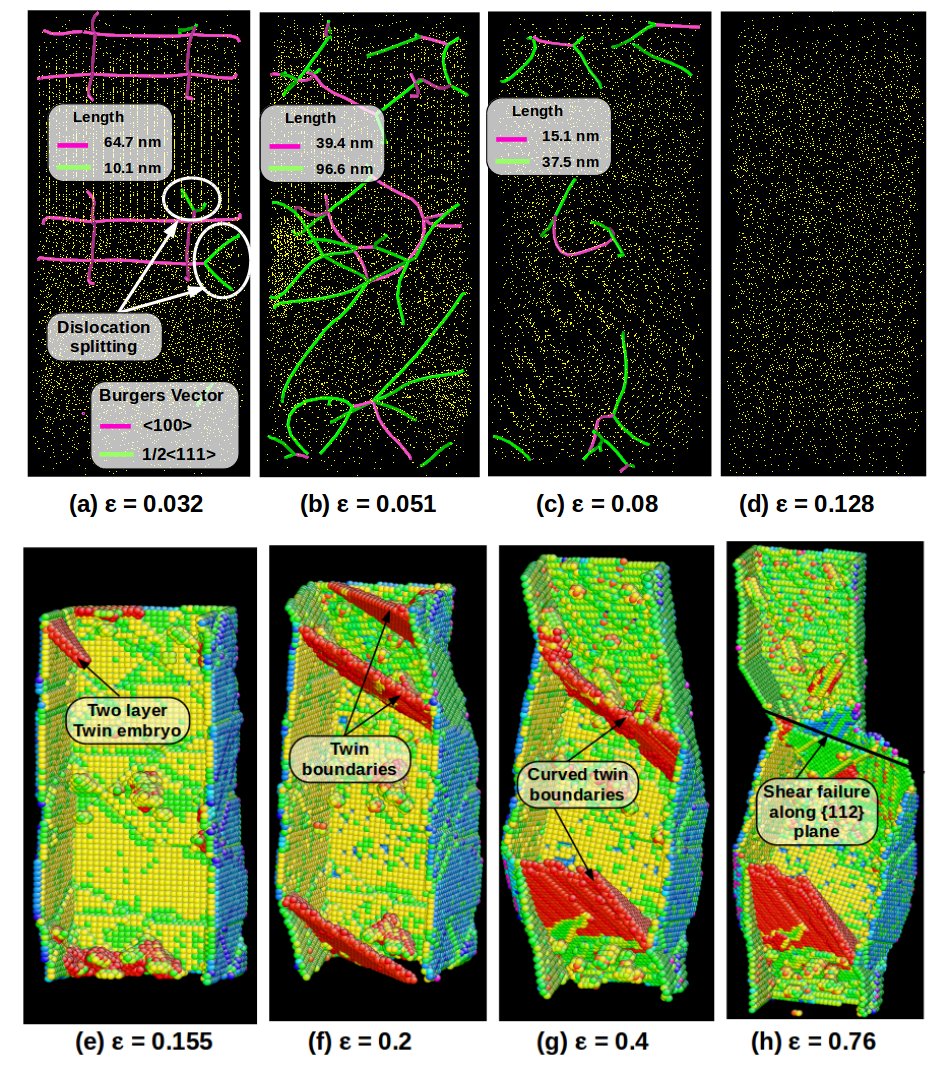}
\caption {\footnotesize Deformation behaviour of $<$100$>$ BCC Fe nanowire containing twist boundary (TWB) 
under tensile loading. In figures (a)-(d), only dislocations are shown using OVITO. In figures (e)–(h), the 
atoms are coloured according to the centro-symmetry parameter. The perfect BCC Fe atoms and surfaces are
removed for clarity.}
\label{twist}
\end{figure}
 
The perfect $<$100$>$ BCC Fe nanowire and the nanowire containing a twist grain boundary display 
significant difference in stress-strain behaviour as well as in operating deformation mechanism. The 
perfect nanowire displays two elastic peaks separated by twinning mode of deformation over large 
plastic strain. In nanowire containing twist boundary, the elastic peaks are separated by dislocation 
slip over small plastic strain. Further, the origin of the second peak is different in these two 
nanowires. In perfect nanowire, it is due to the elastic deformation of the reoriented nanowire, 
while in nanowire containing twist boundary, it is due to elastic deformation of dislocation-free
nanowire. The deformation behaviour of perfect nanowire is in agreement with those reported in 
perfect $<$100$>$ BCC Fe, Mo, and W nanowire \cite{Li,Sai-CMS16,Cao,Mo,Healy,Sai-CMS15}. This 
difference in deformation behaviour arises mainly from the presence of initial dislocation network 
at twist boundary facilitating dislocation slip mechanism at small strains. The continuous splitting 
of $<$100$>$ network sessile dislocations into glissile dislocations and their escape to surface 
leads to dislocation free nanowire close to the perfect one. Following this, twinning takes over 
as the dominant mode of deformation. These results indicate that the presence of initial dislocations 
has dominant effect on twinning mechanism. It has been observed that the deformation by slip of 
full dislocations is preferred over twinning in the presence of dislocations or dislocation networks. 
The study also substantiates the absence of deformation twinning generally observed in bulk materials, 
which inherently contain many dislocations.

\section{Conclusion}

Molecular dynamics simulation results indicate that the presence of \{100\} twist boundary in $<$100$>$ 
BCC Fe nanowire influences the operating deformation twinning mechanism and the associated reorientation 
process. The perfect BCC Fe nanowire deformed by twinning mechanism leading to the reorientation of the 
nanowire, while nanowire containing twist grain boundary deformed by slip at low strains followed by 
twinning at high strains and absence of reorientation. At low strains, deformation by slip of full
dislocations is preferred over twinning in nanowires containing initial dislocations or network 
dislocations. The splitting of $<$100$>$ network sessile dislocations into glissile dislocations and 
their escape to surface leads to dislocation free nanowire, which further deform by twinning at high 
strains. The study also supports the absence of twinning in material containing initial dislocations, 
such as bulk single crystals and bi-crystals with twist boundaries. 

}

\end{document}